\documentclass[aps,twocolumn,amsmath,letterpaper,floatfix]{revtex4}
\usepackage{amssymb}
\usepackage{graphicx}
\usepackage{array}
\usepackage{hhline}
\usepackage{longtable}

\newcommand{\rsp}[1]{\hspace{-0.15em}#1\hspace{-0.15em}}

\begin{document}

\title{Frustrated Further-Neighbor Antiferromagnetic and
Electron-Hopping Interactions in the $d=3$ $tJ$ Model:
Finite-Temperature Global Phase Diagrams from Renormalization-Group
Theory }
\author{C. Nadir Kaplan,$^{1,2,3}$ A. Nihat Berker,$^{4,5,6}$ and Michael
Hinczewski$^{6,7}$} \affiliation{$^1$Department of Physics, Istanbul
Technical University, Maslak 34469, Istanbul, Turkey,}
\affiliation{$^2$Department of Physics, Ko\c{c} University, Sar\i
yer 34450, Istanbul, Turkey,} \affiliation{$^3$Martin Fisher School
of Physics, Brandeis University, Waltham, Massachusetts 02454,
U.S.A.,} \affiliation{$^4$Faculty of Engineering and Natural
Sciences, Sabanc\i~University, Orhanl\i , Tuzla 34956, Istanbul,
Turkey,} \affiliation{$^5$Department of Physics, Massachusetts
Institute of Technology, Cambridge, Massachusetts 02139, U.S.A.,}
\affiliation{$^6$Feza G\"ursey Research Institute, T\"UBITAK -
Bosphorus University, \c{C}engelk\"oy 34684, Istanbul, Turkey}
\affiliation{$^7$Department of Physics, Technical University of
M\"unich, 85748 Garching, Germany}

\begin{abstract}
The renormalization-group theory of the $d=3$ $tJ$ model is extended
to further-neighbor antiferromagnetic or electron-hopping
interactions, including the ranges of frustration.  The global phase
diagram of each model is calculated for the entire ranges of
temperatures, electron densities, further/first-neighbor
interaction-strength ratios.  With the inclusion of further-neighbor
interactions, an extremely rich phase diagram structure is found and
is explained by competing and frustrated interactions.  In addition
to the $\tau_{tJ}$ phase seen in earlier studies of the
nearest-neighbor $d=3$ $tJ$ model, the $\tau_{Hb}$ phase seen before
in the $d=3$ Hubbard model appears both near and away from
half-filling.

PACS numbers: 71.10.Fd, 05.30.Fk, 64.60.De, 74.25.Dw

\end{abstract}

\maketitle
\def\s{\rule{0in}{0.28in}}

\section{Introduction}

The simplest model electron conduction model, including
nearest-neighbor hopping on a lattice and on-site Coulomb repulsion,
is the Hubbard model \cite{Hubbard}. In the limit of very strong
on-site Coulomb repulsion, second-order perturbation theory on the
Hubbard model yields the $tJ$ model \cite{Anderson, BZA}, in which
sites doubly occupied by electrons do not exist.  Studies of the
Hubbard model \cite{HinczewskiBerker1} and of the $tJ$ model
\cite{FalicovBerker}, including spatial anisotropy
\cite{HinczewskiBerker2} and quenched non-magnetic impurities
\cite{HinczewskiBerker3} in good agreement with experiments, have
shown the effectiveness of renormalization-group theory, especially
in calculating phase diagrams at finite temperatures for the entire
range of electron densities in $d=3$.  These calculations have
revealed new phases, dubbed the $\tau$ phases, which occur only in
these electronic conduction models under doping conditions. The
telltale characteristics of the $\tau$ phases are, in contrast to
all other phases of the systems, a non-zero electron-hopping
probability at the largest length scales (at the
renormalization-group thermodynamic-sink fixed points) and the
divergence of the electron-hopping constant $t$ under repeated
rescalings. Furthermore, the phase diagram topologies, the doping
ranges, and the contrasting quantitative $\tau$ and
antiferromagnetic behaviors under quenched impurities
\cite{HinczewskiBerker3} have been in agreement with experimental
findings \cite{Puchkov,Bernhard2}. A benchmark for this
renormalization-group approach has also been established by a
detailed and successful comparison, with the exact numerical results
of the quantum transfer matrix method \cite{Juttner1,Juttner2}, of
the specific heat, charge susceptibility, and magnetic
susceptibility in $d=1$ calculated with our method.\cite{Hinczewski}
Furthermore, results with this method have indicated that no
finite-temperature phase transition occurs in the $tJ$ model in
$d=1$.  A phase separation at zero temperature has been found in
$d=1$ in Ref. \cite{Ogata}.  Thus, the $d=1$ $tJ$ appears to have a
first-order phase transition at zero temperature that disappears as
soon as temperature is raised from zero, as in other $d=1$ models
such as the Ising and Blume-Capel models \cite{Krinsky1,Krinsky2}. A
phase separation \cite{Kivelson,Putikka,Dagotto} occurs in $d=2$ for
low values of $t/J$, but not for $t/J>0.24$.\cite{FalicovBerker} In
$d=3$, a narrow phase separation occurs, as seen in the density -
temperature phase diagrams below. Two distinct $\tau$ phases have
been found in the Hubbard model \cite{HinczewskiBerker1},
$\tau_{Hb}$ and $\tau_{tJ}$, respectively occurring at weak and
strong coupling.  The calculated low-temperature behavior and
critical exponent of the specific heat \cite{HinczewskiBerker1} have
pointed to BCS-like and BEC-like behaviors, respectively. Only the
$\tau_{tJ}$ phase was found in the $tJ$ model.

The current work addresses the issue of whether both $\tau$ phases
can be found in the $tJ$ model, via the inclusion of
further-neighbor antiferromagnetic $(J_2)$ or further-neighbor
electron hopping $(t_2)$ interactions.  We find that, depending on
the temperature and doping level, the further-neighbor interactions
may compete with the further-neighbor effects of the
nearest-neighbor interactions, namely that frustration occurs as a
function of temperature and doping level. This competition (or
reinforcement) between the interactions of successive length scales
underpins the calculated evolution of the phase diagrams.  Global
phase diagrams are obtained for the entire ranges of each type of
further-neighbor interaction. With the inclusion of further-neighbor
interactions, an extremely rich phase diagram structure is found and
is explained by competing and frustrated interactions. Both
$\tau_{Hb}$ and $\tau_{tJ}$ phases are indeed found to occur in the
$tJ$ model with the inclusion of these further-neighbor
interactions.  Furthermore, distinctive lamellar phase diagram
structures of antiferromagnetism interestingly surround the $\tau$
phases in the doped regions.

\section{The $tJ$ Hamiltonian}

On a $d$-dimensional hypercubic lattice, the $tJ$ model is defined
by the Hamiltonian
\begin{multline}
\label{eq:1} -\beta H = P \left[ -t \sum_{\langle ij \rangle,\sigma}
\left(c^\dagger_{i\sigma}c_{j\sigma} +
c^\dagger_{j\sigma}c_{i\sigma}\right)\right.\\
\left. -J \sum_{\langle ij \rangle} \mathbf{S}_i\cdot\mathbf{S}_j +V
\sum_{\langle ij \rangle} n_i n_j + \tilde{\mu}\sum_i n_i \right]
P\,,
\end{multline}
where $\beta=1/k_{B}T$ and, with no loss of generality
\cite{FalicovBerker}, $t\geq 0$ is used. Here $c^\dagger_{i\sigma}$
and $c_{j\sigma}$ are the creation and annihilation operators for an
electron with spin $\sigma=\uparrow$ or $\downarrow$ at lattice site
$i$, obeying anticommutation rules, $n_i = n_{i\uparrow} +
n_{i\downarrow}$ are the number operators where $n_{i\sigma} =
c^\dagger_{i\sigma}c_{i\sigma}$, and $\mathbf{S}_i =
\sum_{\sigma\sigma^\prime} c^\dagger_{i\sigma}
\mathbf{s}_{\sigma\sigma^\prime} c_{i\sigma^\prime}$ is the
single-site spin operator, with $\mathbf{s}$ the vector of Pauli
spin matrices. The projection operator $P = \prod_{i}
(1-n_{i\downarrow}n_{i\uparrow})$ projects out all states with
doubly-occupied sites. The interaction constants $t$, $J$, $V$ and
$\tilde{\mu}$ correspond to electron hopping, nearest-neighbor
antiferromagnetic coupling ($J>0$), nearest-neighbor
electron-electron interaction, and chemical potential, respectively.
From rewriting the $tJ$ Hamiltonian as a sum of pair Hamiltonians
$-\beta H(i,j)$, Eq.~\eqref{eq:1} becomes
\begin{equation}
\label{eq:2} \begin{split} -\beta H = & \sum_{\langle ij \rangle} P
\biggl[ -t \sum_{\sigma} \left(c^\dagger_{i\sigma}c_{j\sigma} +
c^\dagger_{j\sigma}c_{i\sigma}\right)\\ &  - J
\mathbf{S}_i\cdot\mathbf{S}_j + V n_i n_j + \mu ( n_i +n_j)\biggr]
P\\
\equiv &\sum_{\langle ij \rangle} \{-\beta H (i,j)\}\,,
\end{split}
\end{equation}
where $\mu=\tilde{\mu}/2d$. The standard $tJ$ Hamiltonian is a
special case of Eq.~\eqref{eq:2} with $V/J=1/4$, which stems from
second-order perturbation theory on the Hubbard model
\cite{Anderson, BZA}.

\section{Renormalization-Group Transformation}

\subsection{$d=1$ Recursion Relations}

In $d=1$, the Hamiltonian of Eq.~\eqref{eq:2} is
\begin{equation}\label{eq:3}
-\beta H = \sum_i \left\{-\beta H(i,i+1) \right\}\,.
\end{equation}
A decimation eliminates every other one of the successive degrees of
freedom arrayed in a linear chain, with the partition function being
conserved, leading to a length rescaling factor $b=2$. By neglecting
the noncommutativity of the operators beyond three consecutive
lattice sites, a trace over all states of even-numbered sites can be
performed \cite{SuzTak, TakSuz},
\begin{equation}
\label{eq:4} \begin{split} \text{Tr}_{\text{even}} e^{-\beta H}
&=\text{Tr}_{\text{
even}}e^{\sum_{i}\left\{ -\beta H(i,i+1)\right\} }\\
&=\text{Tr}_{\text{even}} e^{\sum_{i}^{\text{
even}}\left\{ -\beta H(i-1,i)-\beta H(i,i+1) \right\} }\\
\simeq \prod_{i}^{\text{even}}\text{Tr}_{i}e&^{\left\{ -\beta
H(i-1,i)-\beta H(i,i+1)\right\} } =\prod_{i}^{\text{
even}}e^{-\beta ^{\prime }H^{\prime }(i-1,i+1)}\\
\simeq& e^{\sum_{i}^{\text{even}}\left\{ -\beta ^{\prime }H^{\prime
}(i-1,i+1)\right\} } =e^{-\beta ^{\prime }H^{\prime }},
\end{split}
\end{equation}
where $-\beta'H'$ is the renormalized Hamiltonian.  This approach,
where the two approximate steps labeled with $\simeq$ are in
opposite directions, has been successful in the detailed solutions
of quantum spin \cite{SuzTak, TakSuz, Tomczak, TomRich1, TomRich2,
KaplanBerker, Sariyer} and electronic \cite{HinczewskiBerker1,
FalicovBerker, HinczewskiBerker2, HinczewskiBerker3} systems.  The
anticommutation rules are correctly accounted within the three-site
segments, at all successive length scales, in the iterations of the
renormalization-group transformation.

The algebraic content of the decimation in Eq.~\eqref{eq:4} is
\begin{equation}
e^{-\beta ^{\prime }H^{\prime }(i,k)}=\mbox{Tr}_{j}e^{-\beta
H(i,j)-\beta H(j,k)}, \label{eq:5}
\end{equation}
where $i,j,k$ are three consecutive sites of the unrenormalized
linear chain. The renormalized Hamiltonian is given by
\begin{equation}\label{eq:17}
\begin{split} -\beta^\prime H^\prime &(i, k) =  P
\biggl[ -t^\prime \sum_{\sigma} \left(c^\dagger_{i\sigma}c_{k\sigma}
+ c^\dagger_{k\sigma}c_{i\sigma}\right)\\ &  - J^\prime
\mathbf{S}_i\cdot\mathbf{S}_k + V^\prime n_i n_k + \mu^\prime ( n_i
+n_k)+G^\prime\biggr] P\,,
\end{split}
\end{equation}
where $G^\prime$ is the additive constant per bond, which is always
generated in renormalization-group transformations, does not affect
the flow of the other interaction constants, and is necessary in the
calculation of expectation values.  The values of the renormalized
(primed) interaction constants appearing in $-\beta'H'$ are given by
the recursion relations extracted from Eq.~\eqref{eq:5}, which will
be given here in closed form, while Appendix A details the
derivation of Eq.~\eqref{eq:6} from Eq.~\eqref{eq:5}:
\begin{gather}
 t'=\frac{1}{2}\ln{\frac{\gamma_4}{\gamma_2}}\,, \quad
 J'=\ln{\frac{\gamma_6}{\gamma_7}}\,,\quad
 V'=\frac{1}{4}\ln{\frac{\gamma_1^4\gamma_6\gamma_7^3}{\gamma_2^4\gamma_4^4}}\,,
 \nonumber\\[5pt]
 \mu'=\mu+\frac{1}{2}\ln{\left(\frac{\gamma_2\gamma_4}{\gamma_1^2}\right)}\,,
 \quad G'=b^dG+\ln{\gamma_1}\,,\label{eq:6}
 \end{gather}

\begin{gather}
\text{where}\qquad\gamma_1=1+2u^3f(\frac{\mu}{2})\,, \nonumber\\[5pt]
 \gamma_2=uf\left(-\frac{\mu}{2}\right)+\frac{1}{2}u^2x^2+\frac{3}{2}u^2vf\left(-\frac{J}{8}+\frac{V}{2}+\frac{\mu}{2}\right)\,,\nonumber\\[5pt]
 \gamma_4=1+\frac{3}{2}u^2v^2+\frac{1}{2}u^2xf\left(\frac{3J}{8}+\frac{V}{2}+\frac{\mu}{2}\right)\,,
 \nonumber\\[5pt]
 \gamma_6=2v^3x+xf\left(-\frac{3J}{8}-\frac{V}{2}-\frac{\mu}{2}\right)\,,\nonumber\\[5pt]
 \gamma_7=\frac{2}{3}vx^3+\frac{4}{3}v^4+vf\left(\frac{J}{8}-\frac{V}{2}-\frac{\mu}{2}\right)\,,\label{eq:8}
\end{gather}

\begin{gather}
\text{and}\qquad v= \exp\left(-J/8+V/2+\mu/2\right)\,,
 \nonumber\\[5pt] x=\exp\left(3J/8+V/2+\mu/2\right)\,,
 \quad u=\exp\left(\mu/2\right)\,,
 \nonumber\\[5pt]
 f(A)=\cosh{\sqrt{2t^2+A^2}}+\frac{A}{\sqrt{2t^2+A^2}}\sinh{\sqrt{2t^2+A^2}}\,.\label{eq:7}
 \end{gather}

\subsection{$d>1$ Recursion Relations}

The Migdal-Kadanoff renormalization-group procedure generalizes our
transformation to $d>1$ through a bond-moving step~\cite{Migdal,
Kadanoff}. Eq.~\eqref{eq:6} can be expressed as a mapping of
interaction constants $\mathbf{K} = \{ G,t,J,V,\mu\}$ onto
renormalized interaction constants, $\mathbf{K}^\prime =
\mathbf{R}(\mathbf{K})$.  The Migdal-Kadanoff procedure strengthens
by a factor of $b^{d-1}$ the bonds of linear decimation, to account
for a bond-moving effect~\cite{Migdal, Kadanoff}.  The resulting
recursion relations for $d>1$ are,
\begin{equation} \label{eq:10}
\mathbf{K^\prime}=b^{d-1} \mathbf{R}(\mathbf{K}),
\end{equation}
which explicitly are
\begin{gather}
 t'=\frac{b^{d-1}}{2}\ln{\frac{\gamma_4}{\gamma_2}},
 J'=b^{d-1}\ln{\frac{\gamma_6}{\gamma_7}},
 V'=\frac{b^{d-1}}{4}\ln{\frac{\gamma_1^4\gamma_6\gamma_7^3}{\gamma_2^4\gamma_4^4}},
 \nonumber\\[5pt]
 \mu'=b^{d-1}\mu+\frac{b^{d-1}}{2}\ln{\left(\frac{\gamma_2\gamma_4}{\gamma_1^2}\right)},
 G'=b^d G+b^{d-1}\ln{\gamma_1}.\label{eq:6a}
 \end{gather}
This approach has been successfully employed in studies of a large
variety of quantum mechanical and classical (\textit{e.g.,}
references in~\cite{HinczewskiBerker1}) systems.

\renewcommand{\arraystretch}{1.3}
\begin{table}[h]
\begin{tabular}{|c|c|c|c|c|}
 \hline
  Phase & \multicolumn{4}{c|}{Interaction constants at sink}\\
  \cline{2-5}
  & $t$   & $\mu$ & $J$ & $V$\\
  \hline
  d (dilute disordered) & 0 & $-\infty$ & 0 & 0\\ \hline
  D (dense disordered) & 0 & $\infty$ & 0 & 0\\ \hline
  AF & 0 & $\infty$ & $-\infty$ & $-\infty$ \\(antiferromagnetic)&&&& $\frac{V}{J}\rightarrow\frac{1}{4} $  \\ \hline
  $\tau_{tJ}$ & $\infty$ & $\infty$ & $\infty$ & $-\infty$ \\
  (BEC-like superconductor)&$\frac{t}{\mu}\rightarrow 1$&& $\frac{J}{\mu}\rightarrow 2$ & $\frac{V}{J}\rightarrow -\frac{3}{4}$\\ \hline
  $\tau_{Hb}$& $-\infty$ & $\infty$ & $-\infty$  & $-\infty$ \\
  (BCS-like superconductor)&$\frac{t}{\mu}\rightarrow -1$&&$\frac{J}{\mu}\rightarrow
  -2$&$\frac{V}{J}\rightarrow \frac{1}{4}$\\ \hline
\end{tabular}
\caption{ Interaction constants at the phase sinks.}
\end{table}

\subsection{Calculation of Phase Diagrams\\
and Expectation Values}

The global flows of Eq.~\eqref{eq:10}, controlled by stable and
unstable fixed points, yield the phase diagrams in temperature
versus chemical potential \cite{Berker0}:  The basin of attraction
of each fixed point corresponds to a single thermodynamic phase or
to a single type of phase transition, according to whether the fixed
point is completely stable (a phase sink) or unstable.  Eigenvalue
analysis of the recursion matrix at an unstable fixed point
determines the order and critical exponents of the phase transitions
at the corresponding basin.

Table I gives the interaction constants {$t, J, V, \mu$} at the $tJ$
model phase sinks.  The $\tau_{tJ}$ and $\tau_{Hb}$ phases are the
only regions where the electron-hopping term $t$ does not
renormalize to zero at the phase sinks.  On the contrary, in these
phases, $t\rightarrow\infty$ and $t\rightarrow -\infty$,
respectively.

To compute temperature versus electron-density (doping) phase
diagrams, thermodynamic densities are calculated by summing along
entire renormalization-group flow trajectories.\cite{McKay} A
density, namely the expectation value of an operator in the
Hamiltonian, is given by
\begin{equation}\label{eq:11}
M_\alpha = \frac{1}{Nd} \frac{\partial \ln Z}{\partial K_\alpha}\,,
\end{equation}
where $K_{\alpha}$ is an element of $\mathbf{K}=\{K_{\alpha}\}$, $Z$
is the partition function, and $N$ is the number of lattice sites.
The recursion relations for densities are
\begin{equation}\label{eq:12}
M_\alpha = b^{-d} \sum_\beta M^\prime_\beta T_{\beta \alpha}\,,
\quad \text{where} \qquad T_{\beta\alpha} \equiv \frac{\partial
K^\prime_\beta}{\partial K_\alpha}\,.
\end{equation}
In terms of the density vector $\mathbf{M}=\{M_{\alpha}\}$ and the
recursion matrix $\mathbf{T}=\{T_{\beta \alpha}\}$,
\begin{equation}\label{eq:13}
\mathbf{T} = \left(\begin{array}{ccccc} b^d & \frac{\partial
G^\prime}{\partial t} & \frac{\partial G^\prime}{\partial J} &
\frac{\partial G^\prime}{\partial V}
& \frac{\partial G^\prime}{\partial \mu}   \\
0  & \frac{\partial t^\prime}{\partial t} & \frac{\partial
t^\prime}{\partial J} & \frac{\partial t^\prime}{\partial V}
& \frac{\partial t^\prime}{\partial \mu}  \\
0 & \frac{\partial J^\prime}{\partial t} & \frac{\partial
J^\prime}{\partial J} & \frac{\partial J^\prime}{\partial V}
& \frac{\partial J^\prime}{\partial \mu}\\
0 & \frac{\partial V^\prime}{\partial t} & \frac{\partial
V^\prime}{\partial J} & \frac{\partial V^\prime}{\partial V}
& \frac{\partial V^\prime}{\partial \mu}\\
0 & \frac{\partial \mu^\prime}{\partial t} &
\frac{\partial\mu^\prime}{\partial J} &
\frac{\partial\mu^\prime}{\partial V} &
\frac{\partial \mu^\prime}{\partial \mu}\\
\end{array}\right),
\end{equation}
Eq.~\eqref{eq:12} simply is
\begin{equation}\label{eq:14}
\mathbf{M} = b^{-d} \mathbf{M}^\prime \cdot \mathbf{T}\,.
\end{equation}
At a fixed point, the density vector $M_\alpha = M_\alpha^\prime
\equiv M_\alpha^\ast$ is the left eigenvector, with eigenvalue
$b^d$, of the fixed-point recursion matrix $\mathbf{T}^\ast$ (Table
II). For non-fixed-points, iterating Eq.~\eqref{eq:14} $n$ times,
\begin{equation}\label{eq:15}
\mathbf{M} = b^{-nd} \mathbf{M}^{(n)}\ \cdot \mathbf{T}^{(n)} \cdot
\mathbf{T}^{(n-1)} \cdot \cdots \cdot \mathbf{T}^{(1)}\,,
\end{equation}
where, for $n$ large enough, the trajectory arrives as close as
desired to a completely stable (phase-sink) fixed point and
$\mathbf{M}^{(n)}\simeq\mathbf{M}^\ast$.  The latter density vector
$\mathbf{M}^\ast$ is the left eigenvector of the recursion matrix
with eigenvalue $b^d$.  When two such density vectors exist, the two
branches of the phase separation of a first-order phase transition
are obtained \cite{McKay,Fisher}, as illustrated with the phase
separations found below.

\renewcommand{\arraystretch}{1.3}
\begin{table}[h]
\begin{tabular}{|c|c|c|c|c|}
 \hline
  Phase sinks & \multicolumn{4}{c|}{Expectation values at sink}\\
  \cline{2-5}
  & $\sum_\sigma \langle c^\dagger_{i\sigma}c_{j\sigma} +
c^\dagger_{j\sigma}c_{i\sigma}\rangle$  & $\langle n_i \rangle$ &
$\langle \mathbf{S}_i \cdot
\mathbf{S}_j \rangle$ & $\langle n_i n_j \rangle$\\
  \hline
  d & 0 & 0 & 0 & 0\\ \hline
  D & 0 & 1 & 0 & 1\\ \hline
  AF & 0 & 1 & $\frac{1}{4}$ & 1 \\\hline
  $\tau_{tJ}$ & $-\frac{2}{3}$ & $\frac{2}{3}$ & $-\frac{1}{4}$ & $\frac{1}{3}$ \\\hline
  $\tau_{Hb}$ & $0.664$& $0.668$ & $0.084$ & $0.336$ \\\hline
\end{tabular}
\caption{Expectation values at the phase sinks.  The expectation
values at a sink epitomize the expectation values throughout its
corresponding phases, because, as explained in Sec. IIIC, the
expectation values at the phase sink underpin the calculation of the
expectation values throughout the corresponding phase which is
constituted from the basin of attraction of the sink.}
\end{table}

\section{Further-Neighbor Interactions, Temperature- and
Doping-Dependent Frustration and Global Phase Diagrams}

For the results presented below, we use the theoretically and
experimentally dictated initial conditions of $V/J=1/4$ and
$t/J=2.25$.

The details of the thermodynamic phases found in this work, listed
in Tables I and II, have been discussed previously within context of
the nearest-neighbor $tJ$ \cite{FalicovBerker, HinczewskiBerker2,
HinczewskiBerker3} and, for the $\tau_{Hb}$ phase, Hubbard
\cite{HinczewskiBerker1} models.  The $\tau_{Hb}$ phase is seen here
in the $tJ$ model with the inclusion of the further-neighbor
antiferromagnetic or electron-hopping interaction. Suffice it to
recall here that the $\tau$ phases are the only phases in which: (1)
the electron-hopping strength does not renormalize to zero, but to
infinity; (2) the electron density does not renormalize to complete
emptiness or complete filling, but to partial emptiness/filling,
leaving room for electron/hole conductivity; (3) the
nearest-neighbor electron occupation probability does not
renormalize to zero or unity, again leaving room for conductivity at
the largest length scales; (4) the electron-hopping expectation
value is non-zero at the largest length scales; (5) the
experimentally observed chemical potential shift as a function of
doping occurs \cite{HinczewskiBerker2}; and (6) a low level $(\sim
6\%)$ of quenched non-magnetic impurities causes total
disappearance, in contrast to the antiferromagnetic phase $(\sim
40\%$ for total disappearance) \cite{HinczewskiBerker3}, again as
seen experimentally.  The low-temperature behavior and critical
exponent of the specific heat \cite{HinczewskiBerker1} have pointed
to BCS-like and BEC-like behaviors for the $\tau_{Hb}$ and
$\tau_{tJ}$ phases, respectively.

The only approximations in obtaining the results below are the
Suzuki-Takano and Migdal-Kadanoff procedures, explained above in
Secs. IIIA and IIIB respectively. There are no further assumptions
in Secs.IVA and IVB below.

\subsection{The $t_2$ Model}

\begin{figure}[!t]
\vspace{1em} \centering \includegraphics*[scale=0.7]{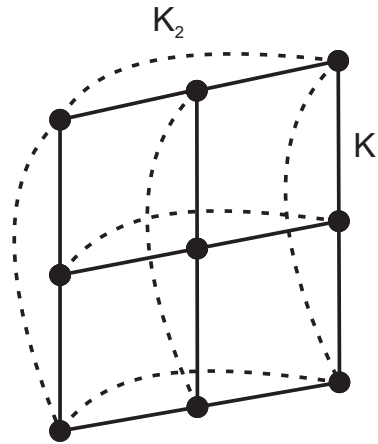}
\caption{Construction of the further-neighbor models.  Part of a
single plane of the three-dimensional model studied here is shown.}
\end{figure}

\begin{figure*}
\centering
\includegraphics*[width=1\textwidth]{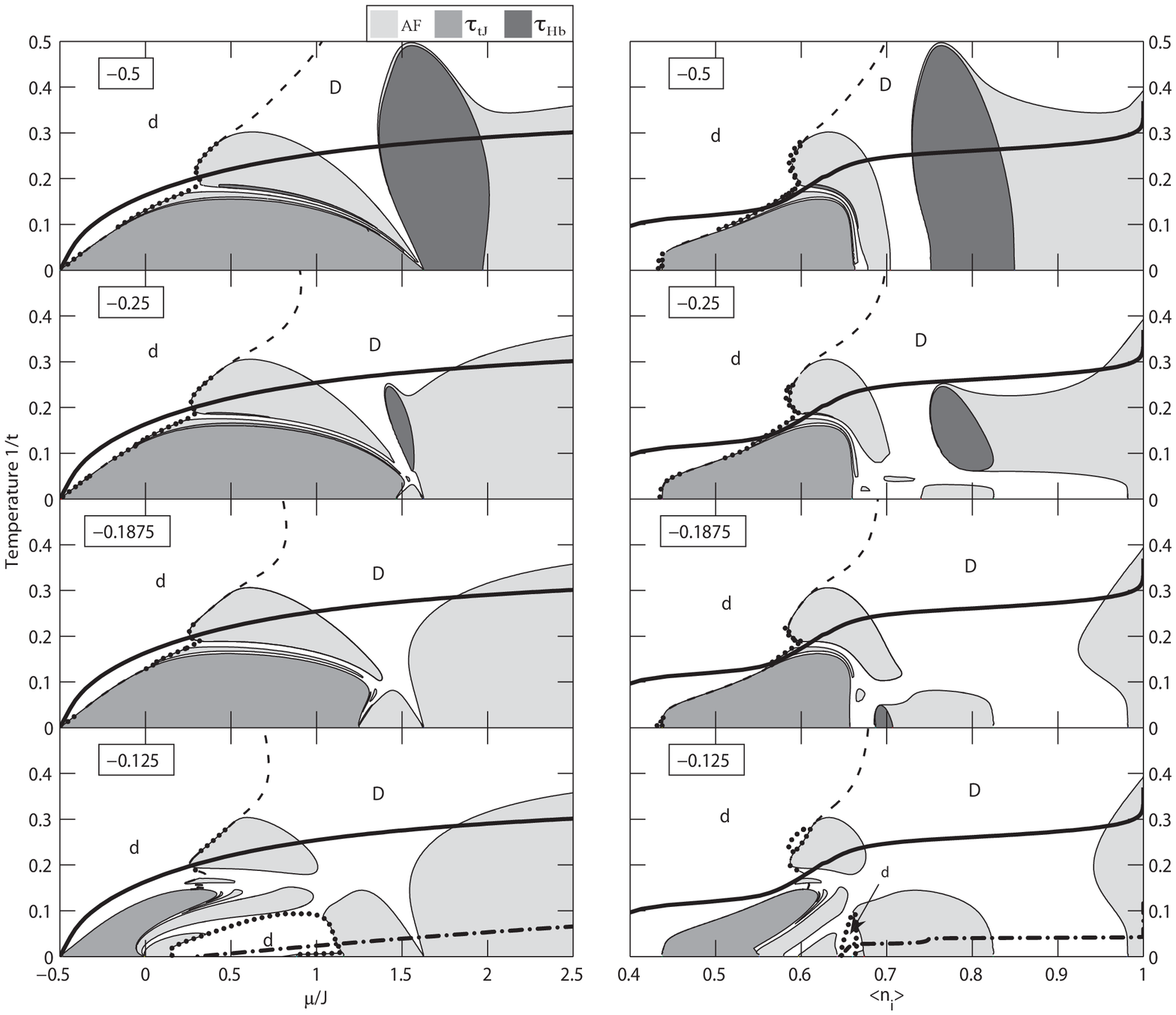}
\caption{Global phase diagram of the further-neighbor $t_{2}$ model
for $t/J=2.25$ in temperature vs. chemical potential (first column)
and, correspondingly, temperature versus electron density (second
column).  The relation $t/J=2.25$ is used for all
renormalization-group trajectory initial conditions.  The $t_2/t$
values are given in boxes.  The dilute disordered (d), dense
disordered (D), antiferromagnetic AF (lightly colored), $\tau_{tJ}$
(medium colored), and $\tau_{Hb}$ (darkly colored) phases are seen.
Second-order phase transitions are drawn with full curves,
first-order transitions with dotted curves. Phase separation occurs
between the dense (D) and dilute (d) disordered phases, in the
unmarked areas within the dotted curves in the electron density vs.
temperature diagrams.  These areas are bounded, on the right and and
the left, by the two branches of phase separation densities,
evaluated by renormalization-group theory as explained in Sec.IIIC.
Note that these coexistence regions between dense (D) and dilute (d)
disordered phases are very narrow.  Dashed curves are not phase
transitions, but disorder lines between the dense and dilute
disordered phases.  As explained in the text, on each side of the
thick full curves (not a phase boundary), the further-neighbor
electron hopping affects the $\tau$ phases oppositely.  On the
dash-dotted curve (also not a phase boundary; overlaps, for
$t_2/t=0$, with the thick full curve) electron hopping in the system
is frustrated.}
\end{figure*}

\begin{figure*}
\centering
\includegraphics*[width=1\textwidth]{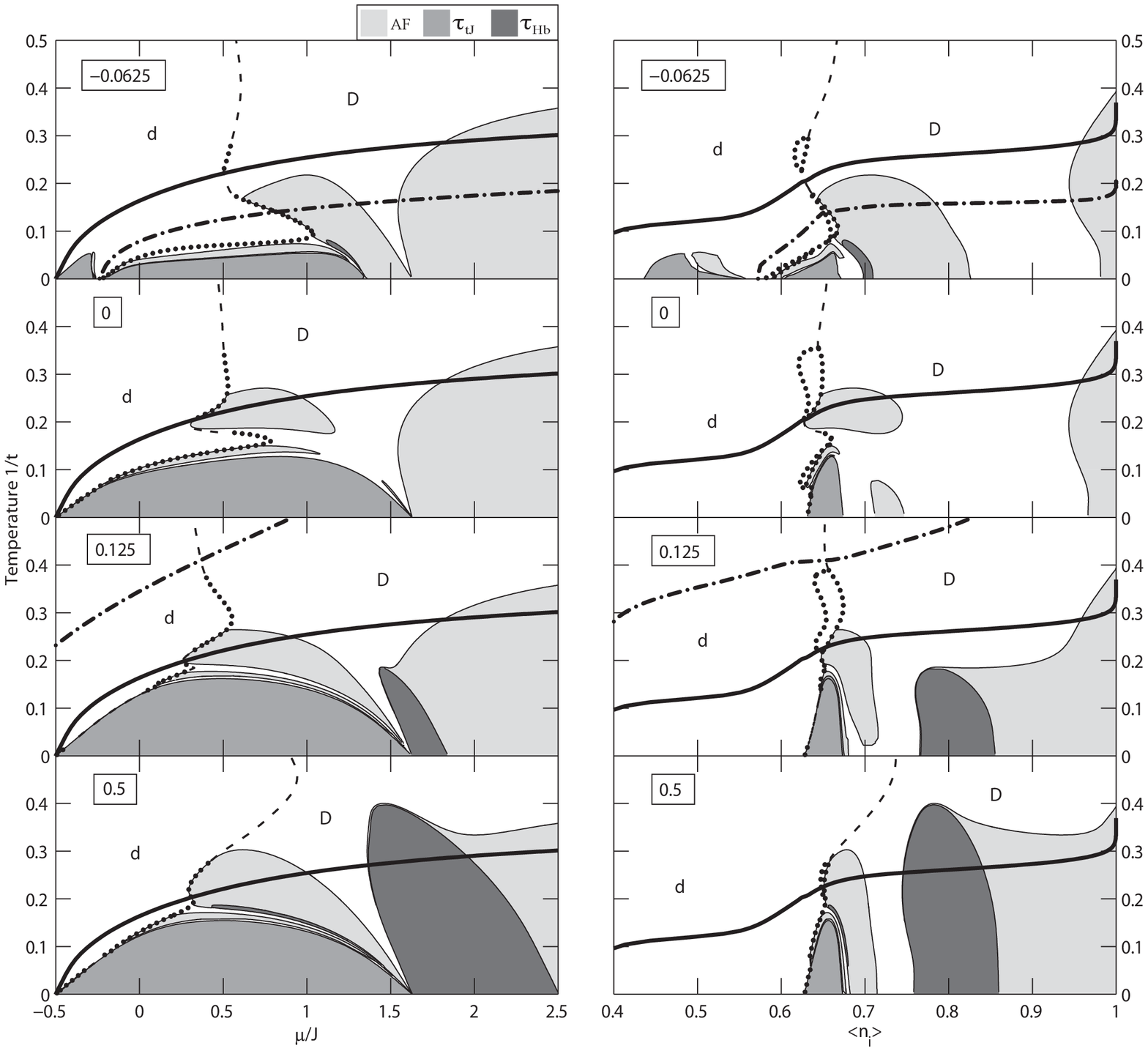}
\caption{The continuation of the global phase diagram in Fig.~2.}
\end{figure*}

\begin{figure*}
\centering
\includegraphics*[width=1\textwidth]{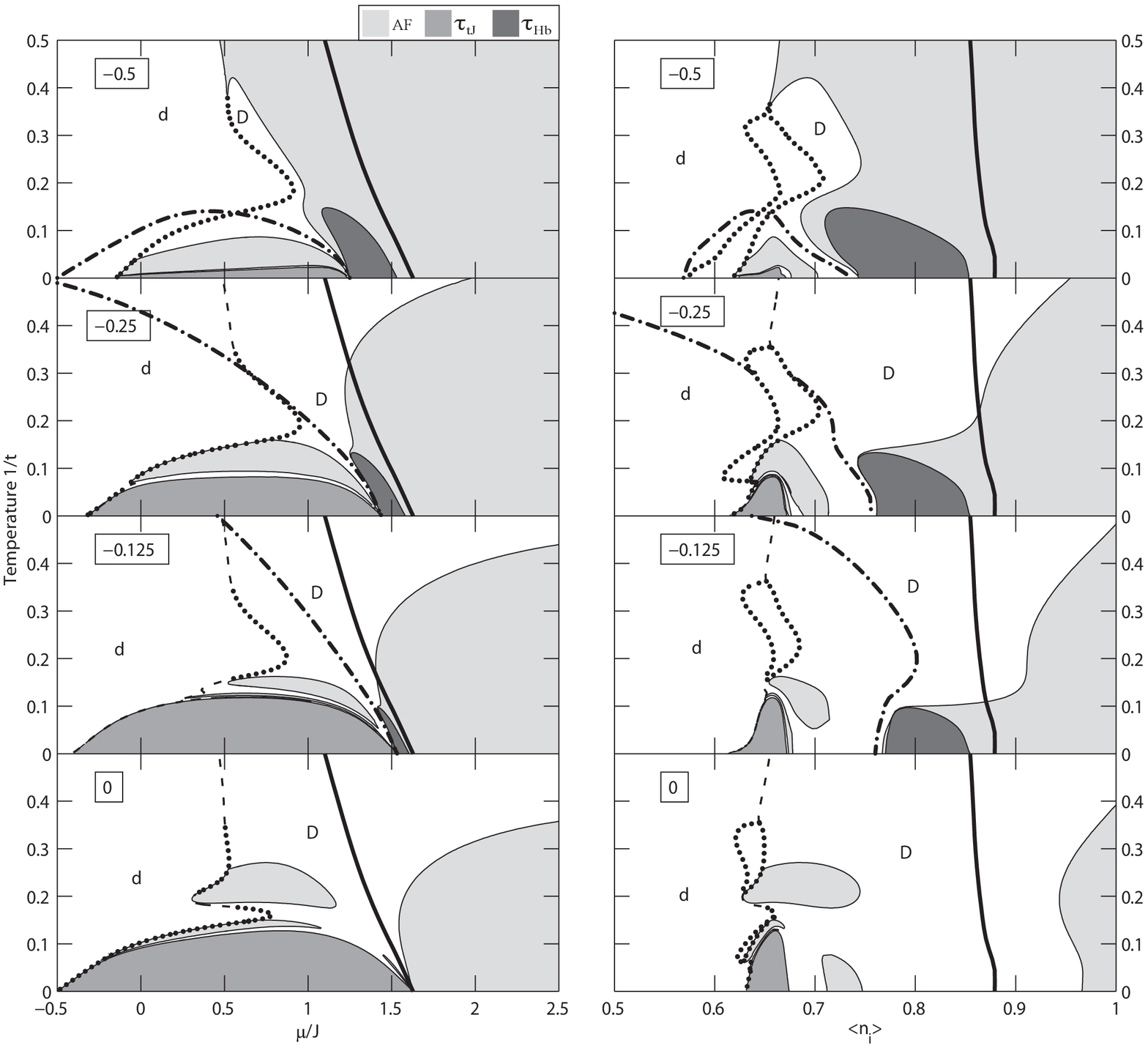}
\caption{Global phase diagrams of the further-neighbor $J_{2}$ model
for $t/J=2.25$ in temperature vs. chemical potential (first column)
and, correspondingly, temperature versus electron density (second
column).  The relation $t/J=2.25$ is used for all
renormalization-group trajectory initial conditions.  The $J_2/J$
values are given in boxes. The dilute disordered (d), dense
disordered (D), antiferromagnetic AF (lightly colored), $\tau_{tJ}$
(medium colored), and $\tau_{Hb}$ (darkly colored) phases are seen.
Second-order phase transitions are drawn with full curves,
first-order transitions with dotted curves. Phase separation occurs
between the dense (D) and dilute (d) disordered phases, in the
unmarked areas within the dotted curves in the electron density vs.
temperature diagrams.  These areas are bounded, on the right and and
the left, by the two branches of phase separation densities,
evaluated by renormalization-group theory as explained in Sec.IIIC.
Note that these coexistence regions between dense (D) and dilute (d)
disordered phases are very narrow.  Dashed curves are not phase
transitions, but disorder lines between the dense and dilute
disordered phases.  As explained in the text, on each side of the
thick full curves (not a phase boundary), the further-neighbor
interaction affects the antiferromagnetic phase oppositely. On the
dash-dotted curve (also not a phase boundary; overlaps, for
$J_2/J=0$, with the thick full curve), antiferromagnetism in the
system is frustrated.}
\end{figure*}

\begin{figure*}
\centering
\includegraphics*[width=1\textwidth]{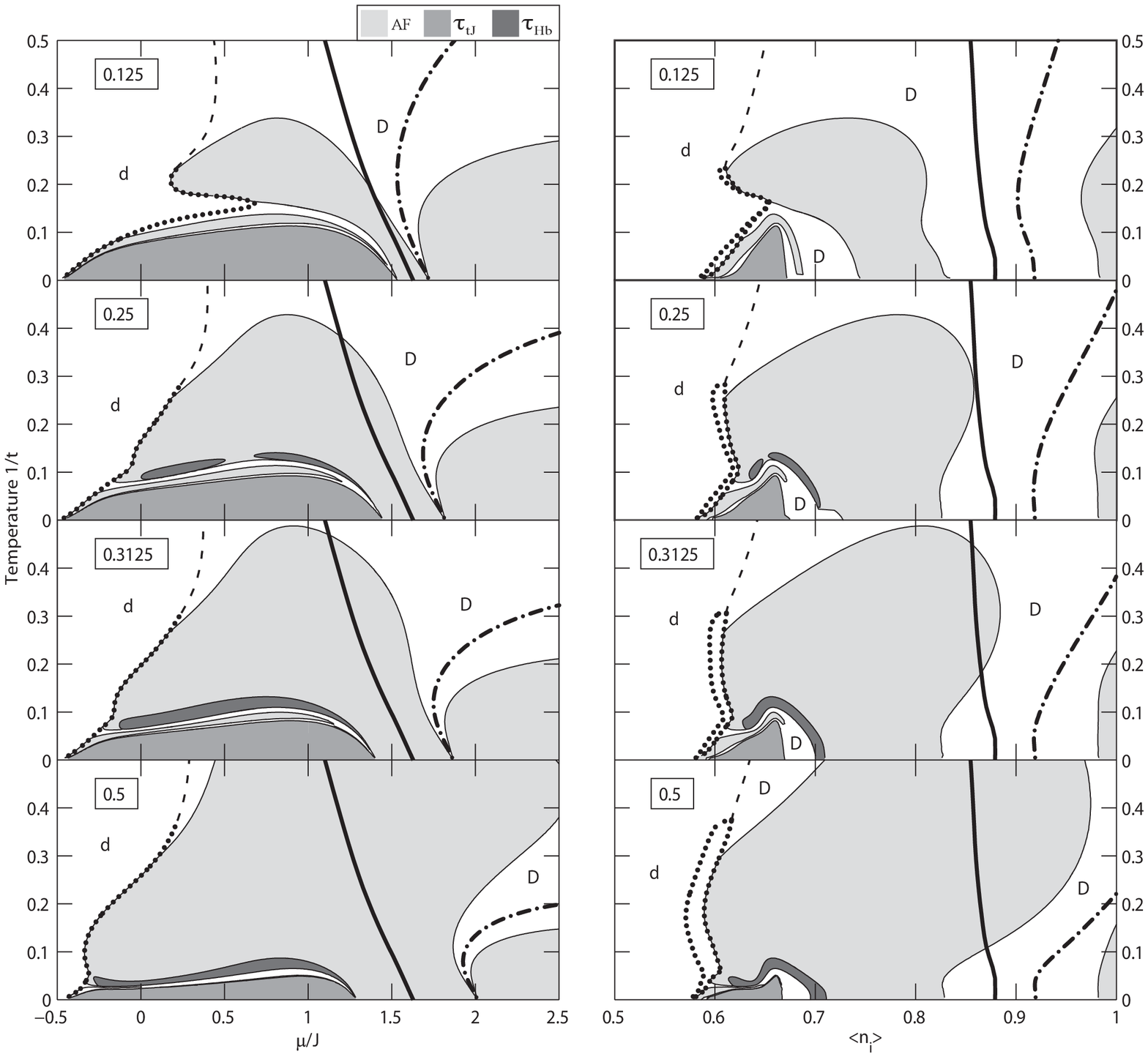}
\caption{The continuation of the global phase diagrams in Fig.~4.}
\end{figure*}

The $t_2$ model includes further-neighbor electron-hopping
interaction, as shown in Fig.~1.  The three-site Hamiltonian,
between the lattice nodes at the lowest length scale, has the form:
\begin{equation}
\label{eq:16}
\begin{split}
-\beta H(i,j,k) = & -\beta H(i,j)-\beta H(j,k)\\ &-t_{2}
\sum_{\sigma} \left(c^\dagger_{i\sigma}c_{k\sigma} +
c^\dagger_{k\sigma}c_{i\sigma}\right) \,,
\end{split}
\end{equation}
where $-\beta H(i,j)$ is given in Eq.~\eqref{eq:2}, so that the
first equation of Eq.~\eqref{eq:6} gets modified as
\begin{equation}
 t'=\frac{1}{2}\ln{\frac{\gamma_4}{\gamma_2}} + t_2\,,\label{eq:16b}
\end{equation}
only for the first renormalization.  Thus, for $d=3$, the first
equation of Eq.~\eqref{eq:6a} gets modified as
\begin{equation}
t'=2\,\ln{\frac{\gamma_4}{\gamma_2}} + 4\,t_2\,,\label{eq:16bb}
\end{equation}
only for the first renormalization. Thus, the hopping-strength $t_2$
contributes to the first renormalization, but is not regenerated by
this first renormalization.  Note that the quantitative memory of
the further-neighbor interaction is kept in all subsequent
renormalization-group steps, as the flows are modified by the
different values of the first-renormalized interactions due to the
effect of the further-neighbor interaction.  The subsequent global
renormalization-group flows are in the space of  $t,J,V,\mu$, as is
the case in position-space renormalization-group treatments
\cite{Berker,Caflisch1,Caflisch2} using a prefacing transformation.
Which surfaces in this large (4-dimensional) flow space of
$t,J,V,\mu$ are accessed is controlled by the original
further-neighbor interaction.  Thus, the further-neighbor
interaction $t_2$ shifts the value of $t'$ obtained after the first
renormalization-group transformation, as dictated by the physical
model (Fig.1). Since the value of the first-renormalized $t'$ in the
absence of $t_2$ already has a complicated dependence on the
unrenormalized temperature and electron density, the variety of
phase diagrams is obtained. Indeed, the effect of the
further-neighbor interaction is dependent on the electron density,
temperature, and other interactions in the system, due to the
presence of the first-term in Eq.~\eqref{eq:16bb}, which is the key
to the resulting spectacularly different phase diagrams as the
further-neighbor interaction is varied.  (1) If the two terms in
Eq.~\eqref{eq:16b} are of the same sign, the nearest-neighbor and
further-neighbor electron hopping terms of the original system
reinforce each other and the $\tau$ phases are enhanced.  (2) If the
two terms are of opposite signs, the nearest-neighbor and
further-neighbor electron hopping terms of the original system
compete with each other and, with the introduction of
further-neighbor electron hopping, the $\tau$ phases are initially
suppressed, but enhanced as further-neighbor hopping becomes
dominant.  The two regimes (1) and (2) are separated by the thick
full lines in the phase diagrams in Figs. 2 and 3.  In the case (2)
of opposite signs, when the two terms cancel out each other, the
system is frustrated, in which case, after the first
renormalization, there is no electron hopping in the system. Since
this condition is closed under renormalization, both on physical
grounds and of course in our recursion relations (Eq.~\eqref{eq:6}),
no $\tau$ phase can occur in such a system.  The dash-dotted curves
in Figs. 2 and 3 indeed show such systems.  These competition,
reinforcement, and frustration effects are temperature and doping
dependent. These, and all other physical effects, do not depend on
the sign of nearest-neighbor $t$ of the original unrenormalized
system, due to the symmetry of hypercubic lattices
\cite{FalicovBerker} and as seen in Eq.~\eqref{eq:7}.

Figs. 2 and 3 give the global phase diagram of the $t_2$ model, as a
function of temperature, electron density, chemical potential, and
$t_2/t$. The values of the hopping-strength ratios $t_2/t$ for the
consecutive panels in these figures are chosen so that they
sequentially produce the qualitatively different phase-diagram
cross-sections, thereby revealing the evolution in the global phase
diagram.  Second-order phase transitions are drawn with full curves,
first-order transitions with dotted curves. Phase separation occurs
between the dense (D) and dilute (d) disordered phases, in the
unmarked areas within the dotted curves in the electron density vs.
temperature diagrams.  These areas are bounded, on the right and and
the left, by the two branches of phase separation densities,
evaluated by renormalization-group theory as explained in Sec.IIIC.
Note that these coexistence regions between dense (D) and dilute (d)
disordered phases are very narrow.

The cross-section $t_2=0$ is the phase diagram obtained in previous
work \cite{FalicovBerker}. This diagram contains the $\tau_{tJ}$
phase between $33-37\%$ hole doping away from half-filling and at
temperatures $1/t<0.12$.  The thick full curve here gives the
systems devoid of electron hopping due to the combined effects of
temperature and doping on a nearest-neighbor-only interaction
system. The first term of Eq.~\eqref{eq:16b} is positive on the high
density/chemical potential, low temperature side of the thick full
curve and negative on the low chemical potential/density, high
temperature side of the thick full curve. Thus, the inclusion of
$t_2>0$ will create competition and frustration (respectively
reducing and eliminating the $\tau$ phases) on the low chemical
potential/density, high temperature side of the curve discussed
here, reinforcement (enhancing the $\tau$ phases) on the high
chemical potential/density, low temperature side of the same curve.
The opposite occurs at $t_2<0$. The thick full (no-hopping) curve of
$t_2=0$ is included, again as thick and full, in the $t_2\ne 0$
phase diagrams and the effects discussed here are seen in the
evolution, in both directions, of the global phase diagram.

Pursuing the negative values of $t_2$, we see at $t_{2}/t=-0.0625$
that the $\tau_{tJ}$ phase, being below the thick full curve, is
indeed reduced and bisected into two disconnected regions by the
frustration (dash-dotted) curve.  At the more negative value of
$t_{2}/t$ = -0.125, only the higher doping region of the $\tau_{tJ}$
phase remains and is enhanced as explained after Eq.~\eqref{eq:16b},
extending through a wider range to $45-55\%$ hole doping. The
antiferromagnetic and disordered phases take part in a complex
lamellar structure, in a narrow band between $35-45\%$ hole doping
at low temperatures.  At the even more negative values of
$t_{2}/t=-0.25$ and $-0.5$, the $\tau_{tJ}$ phase appears in a wide
range of hole doping, between $35-55\%$. Besides the complex
lamellar structure of antiferromagnetic and disordered phases, we
also see that the $\tau_{Hb}$ phase participates in the lamellar
phase structure and, separately, appears adjacently to the
antiferromagnetic phase near half-filling. Particularly near
half-filling, the $\tau_{Hb}$ phase which evolves adjacently to the
antiferromagnetic phase reaches to the higher temperatures of
$1/t\sim0.5$.  This topology is identical to that obtained for the
Hubbard model \cite{HinczewskiBerker1}.

For the positive values of $t_{2}/t$, the $\tau$ phases are enhanced
as explained after Eq.~\eqref{eq:16b} and the topology quickly
evolves to that encountered in the Hubbard model. The $\tau_{tJ}$ is
not bisected by the frustration (dash-dotted) curve and appears
between $33-37\%$ hole doping as a continuation of the structure at
$t_2=0$. The $\tau_{Hb}$ phase occurs again in two distinct regions
and the one which lies nearer to half-filling again extends to high
temperatures.

\subsection{The $J_2$ Model}

The $J_2$ model includes further-neighbor antiferromagnetic
interaction, as shown in Fig.~1.  The three-site Hamiltonian,
between the lattice nodes at the lowest length scale, has the form:
\begin{equation}
\label{eq:16c}
\begin{split}
-\beta H(i,j,k) = & -\beta H(i,j)-\beta H(j,k)\\ &-J_{2}
\sum_{\langle ik \rangle} \mathbf{S}_i\cdot\mathbf{S}_k \,,
\end{split}
\end{equation}
where $-\beta H(i,j)$ is given in Eq.~\eqref{eq:2}, so that the
second equation of Eq.~\eqref{eq:6} gets modified as
\begin{equation}
 J'=\ln{\frac{\gamma_6}{\gamma_7}} + J_2\,,\label{eq:16d}
\end{equation}
only for the first renormalization. Thus, for $d=3$, the second
equation of Eq.~\eqref{eq:6a} gets modified as
\begin{equation}
 J'=4\,\ln{\frac{\gamma_6}{\gamma_7}} + 4\,J_2\,,\label{eq:16d}
\end{equation}
only for the first renormalization. Again, the interaction $J_2$
contributes to the first renormalization, but is not regenerated by
this first renormalization.  Reinforcement or competition occurs
when $J_2$ is, respectively, of same or opposite sign as the first
term in Eq.~\eqref{eq:16d}.  These two regimes are again separated
by the thick full lines in the phase diagrams of Figs. 3 and 4,
while again frustration occurs on the dash-dotted lines.  In the
reinforcement regime, we expect a large extent of the
antiferromagnetic phase. The $\tau_{Hb}$ phase is also expected to
grow in the reinforced region, for it is found along the temperature
extent of the antiferromagnetic phase.

Figs. 4 and 5 show the global phase diagram of the $J_2$ model, as a
function of temperature, electron density, chemical potential, and
$J_2/J$.  Again, the values of the coupling-strength ratios $J_2/J$
for the consecutive panels in these figures are chosen so that they
sequentially produce the qualitatively different phase-diagram
cross-sections, thereby revealing the evolution in the global phase
diagram.  Again, the phase separation regions of the first-order
phase transitions are very narrow.  For negative values of $J_2/J$,
the antiferromagnetic phase is enhanced, both near half-filling by
the mechanism explained after Eq.~\eqref{eq:16d} and, separately and
to a lesser extent, displacing the $\tau_{tJ}$ phase. The latter
behavior is similar to that seen under the introduction of quenched
impurities, both experimentally \cite{Julien, Watanabe, Itoh} and
from renormalization-group theory \cite{HinczewskiBerker3}. The
$\tau_{Hb}$ phase improves near the large antiferromagnetic region
near half-filling. At $J_2/J=-2$, the $\tau_{Hb}$ phase is found in
a wide range of hole doping, namely between $15-30\%$. Another
interesting result is that the $\tau_{tJ}$ phase is depressed in
temperature but remains stable in the interval of $33-37\%$ hole
doping.

For positive values of $J_2/J$, the antiferromagnetic phase is
reduced in the region near half-filling and enhanced in the region
near the $\tau_{tJ}$ phase, for reasons explained after
Eq.~\eqref{eq:16d}.  The $\tau_{Hb}$ phase grows adjacently to the
enhanced antiferromagnetic region, being located above the
$\tau_{tJ}$ phase, causing a complex structure at higher hole
dopings and low temperatures.

\subsection{Conclusion}

We have shown that the $tJ$ model with further-neighbor
antiferromagnetic $(J_2)$ or further-neighbor electron hopping
$(t_2)$ interactions exhibits extremely rich global phase diagrams.
The phase separation regions of the first-order phase transions are
very narrow.  Furthermore, these calculated phase diagrams are
understood in terms of the competition and frustration of nearest-
and further-neighbor interactions.  We find that the two types of
$\tau$ phases, previously seen in the Hubbard model, occur in the
$tJ$ model with the inclusion of further-neighbor interactions.

\begin{acknowledgments}
This research was supported by the Scientific and Technological
Research Council of Turkey (T\"UB\.ITAK) and by the Academy of
Sciences of Turkey.
\end{acknowledgments}

\begin{table}[h]
\begin{tabular}{|c|c|c|c|c|}
 \hline
  $n$ & $p$ & $s$ & $m_s$ & Two-site eigenstates\\
  \hline
  $0$ & $+$ & $0$ & $0$ &$|\phi_{1}\rangle=|\circ\circ\rangle$ \\
  \hline
  $1$ & $+$ & $1/2$ & $1/2$ &$|\phi_{2}\rangle=\frac{1}{\sqrt{2}}\{|\uparrow
  \circ\rangle+|\circ\uparrow\rangle\}$\\ \hline
  $1$ & $-$ & $1/2$ & $1/2$ &$|\phi_{4}\rangle=\frac{1}{\sqrt{2}}\{|\uparrow
  \circ\rangle-|\circ\uparrow\rangle\}$\\ \hline
  $2$ & $-$ & $0$ & $0$ & $|\phi_{6}\rangle=\frac{1}{\sqrt{2}}\{|\uparrow\downarrow\rangle
  -|\downarrow\uparrow\rangle\}$\\ \hline
  $2$ & $+$ & $1$ & $1$ &
  $|\phi_{7}\rangle=|\uparrow\uparrow\rangle$\\ \hline
  $2$ & $+$ & $1$ & $0$ &
  $|\phi_{9}\rangle=\frac{1}{\sqrt{2}}\{|\uparrow\downarrow\rangle+|\downarrow
  \uparrow\rangle\}$\\ \hline
\end{tabular}
\caption{The two-site basis states, with the corresponding particle
number ($n$), parity ($p$), total spin ($s$), and total spin
$z$-component ($m_s$) quantum numbers.  The states
$|\phi_{3}\rangle$, $|\phi_{5}\rangle$, and $|\phi_{8}\rangle$ are
obtained by spin reversal from $|\phi_{2}\rangle$,
$|\phi_{4}\rangle$, and $|\phi_{7}\rangle$, respectively.}
\end{table}

\begin{table}
\begin{tabular}{|c|c|c|c|c|}
 \hline
  $n$ & $p$ & $s$ & $m_s$ & Three-site eigenstates\\
  \hline
  $0$ & $+$ & $0$ & $0$ &$|\psi_{1}\rangle=|\circ\circ\,\circ\rangle$ \\
  \hline
  $1$ & $+$ & $1/2$ & $1/2$ &$|\psi_{2}\rangle=|\circ
  \uparrow
  \circ\rangle,\: |\psi_{3}\rangle=\frac{1}{\sqrt{2}}\{|\uparrow
  \circ\,\circ\rangle+|\circ\,\circ\uparrow\rangle\}$\\ \hline
  $1$ & $-$ & $1/2$ & $1/2$ &$|\psi_{6}\rangle=\frac{1}{\sqrt{2}}\{|\uparrow
  \circ\,\circ\rangle-|\circ\,\circ\uparrow\rangle\}$\\ \hline
$2$ & $+$ & $0$ & $0$ &
  $|\psi_{8}\rangle=\frac{1}{2}\{|\uparrow\downarrow\circ\rangle-
  |\downarrow\uparrow\circ\rangle-|\circ\uparrow\downarrow\rangle+
  |\circ\downarrow\uparrow\rangle\}$\\ \hline
   $2$ & $-$ & $0$ & $0$ &
  $|\psi_{9}\rangle=\frac{1}{2}\{|\uparrow\downarrow\circ\rangle-
  |\downarrow\uparrow\circ\rangle+|\circ\uparrow\downarrow\rangle-
  |\circ\downarrow\uparrow\rangle\},$\\
  &&&&$|\psi_{10}\rangle=\frac{1}{\sqrt{2}}\{|\uparrow\circ\downarrow\rangle-|\downarrow\circ\uparrow
  \rangle\}$\\\hline
  $2$ & $+$ & $1$ & $1$ &
  $|\psi_{11}\rangle=|\uparrow\circ\uparrow\rangle,\:
  |\psi_{12}\rangle=\frac{1}{\sqrt{2}}\{|\uparrow\uparrow\circ\rangle+|\circ\uparrow\uparrow
  \rangle\}$\\ \hline
  $2$ & $+$ & $1$ & $0$ &
  $|\psi_{13}\rangle=\frac{1}{2}\{|\uparrow\downarrow\circ\rangle+
  |\downarrow\uparrow\circ\rangle+|\circ\uparrow\downarrow\rangle+
  |\circ\downarrow\uparrow\rangle\},$\\
  &&&& $|\psi_{14}\rangle=\frac{1}{\sqrt{2}}
  \{|\uparrow\circ\downarrow\rangle+|\downarrow\circ\uparrow
  \rangle\}$\\ \hline
  $2$ & $-$ & $1$ & $1$ &
  $|\psi_{17}\rangle=\frac{1}{\sqrt{2}}\{|\uparrow\uparrow\circ\rangle-|\circ\uparrow\uparrow
  \rangle\}$\\ \hline
  $2$ & $-$ & $1$ & $0$ &
  $|\psi_{18}\rangle=\frac{1}{2}\{|\uparrow\downarrow\circ\rangle+
  |\downarrow\uparrow\circ\rangle-|\circ\uparrow\downarrow\rangle-
  |\circ\downarrow\uparrow\rangle\}$\\ \hline
  $3$ & $+$ & $1/2$ & $1/2$ &
  $|\psi_{20}\rangle=\frac{1}{\sqrt{6}}\{2|\uparrow\downarrow\uparrow\rangle-|\uparrow\uparrow
  \downarrow\rangle-|\downarrow\uparrow\uparrow\rangle\}$\\
  \hline
  $3$ & $-$ & $1/2$ & $1/2$ &
  $|\psi_{22}\rangle=\frac{1}{\sqrt{2}}\{|\uparrow\uparrow
  \downarrow\rangle-|\downarrow\uparrow\uparrow\rangle\}$\\
  \hline
  $3$ & $+$ & $3/2$ & $3/2$ &
  $|\psi_{24}\rangle=|\uparrow\uparrow\uparrow\rangle$ \\ \hline
  $3$ & $+$ & $3/2$ & $1/2$ &
  $|\psi_{25}\rangle=\frac{1}{\sqrt{3}}\{|\uparrow\downarrow\uparrow\rangle+|\uparrow\uparrow
  \downarrow\rangle+|\downarrow\uparrow\uparrow\rangle\}$ \\ \hline
\end{tabular}
\caption{The three-site basis states, with the corresponding
particle number ($n$), parity ($p$), total spin ($s$), and total
spin $z$-component ($m_s$) quantum numbers. The states
$|\psi_{4-5}\rangle$, $|\psi_{7}\rangle$, $|\psi_{15-16}\rangle$,
$|\psi_{19}\rangle$, $|\psi_{21}\rangle$, $|\psi_{23}\rangle$,
$|\psi_{26-27}\rangle$ are obtained by spin reversal from
$|\psi_{2-3}\rangle$, $|\psi_{6}\rangle$, $|\psi_{11-12}\rangle$,
$|\psi_{17}\rangle$, $|\psi_{20}\rangle$, $|\psi_{22}\rangle$,
$|\psi_{24-25}\rangle$, respectively.}
\end{table}

\begingroup
\squeezetable
\begin{table}
\begin{gather*}
\begin{array}{|c||c|c|c|c|c|c|} \hline & \parbox{0.08in}{$\phi_{1}$} &\phi_{2} &\phi_{4} & \phi_{6} &
\phi_{7} &\phi_{9}\\
\hhline{|=#=|=|=|=|=|=|} \parbox{0.1in}{$\phi_{1}$} & \parbox{0.08in}{$G^\prime$} & \multicolumn{5}{c|}{}\\
\cline{1-3} \parbox{0.1in}{$\phi_{2}$} && \parbox{0.5in}{\centering $-t^\prime+\mu^\prime+G^\prime$} & \multicolumn{4}{c|}{0}\\
\cline{1-1}\cline{3-4} \parbox{0.1in}{$\phi_{4}$} &
\multicolumn{2}{c|}{}
&\parbox{0.5in}{\centering $t^\prime + \mu^\prime+G^\prime$} &\multicolumn{3}{c|}{}\\
\cline{1-1}\cline{4-5} \parbox{0.1in}{$\phi_{6}$} &
\multicolumn{3}{c|}{} &
\multicolumn{1}{c|}{\parbox{0.575in}{\centering $\frac{3}{4}J^\prime
+V^\prime+2\mu^\prime+G^\prime$}} &
\multicolumn{2}{c|}{}\\
\cline{1-1}\cline{5-6} \parbox{0.1in}{$\phi_{7}$} &
\multicolumn{4}{c|}{0}
&\multicolumn{1}{c|}{\parbox{0.485in}{\centering $-\frac{1}{4}J^\prime+V^\prime+2\mu^\prime+G^\prime$}}&\\
\cline{1-1}\cline{6-7} \parbox{0.1in}{$\phi_{9}$} &
\multicolumn{5}{c|}{}
& \parbox{0.485in}{\centering $-\frac{1}{4}J^\prime+V^\prime+2\mu^\prime+G^\prime$}\\
\hline
\end{array}
\end{gather*}
\caption{Block-diagonal matrix of the renormalized two-site
Hamiltonian $-\beta^\prime H^\prime(i,k)$.  The Hamiltonian being
invariant under spin-reversal, the spin-flipped matrix elements are
not shown.}
\end{table}
\endgroup

\begingroup
\squeezetable
\begin{table}
\begin{gather*}
\begin{array}{|c||c|}\hline
 & \psi_{1}\\
\hhline{|=#=|} \psi_{1} & 0\\ \hline
\end{array}\\
\begin{array}{|c||c|c|}\hline
 & \psi_{2} & \psi_{3}\\
\hhline{|=#=|=|} \psi_{2} & 2\mu & -\sqrt{2}t\\
\hline \psi_{3} & -\sqrt{2}t& \mu\\
\hline
\end{array}\quad
\begin{array}{|c||c|c|}\hline
 & \psi_{6} &\psi_{8}\\
\hhline{|=#=|=|} \psi_{6} &\mu & 0\\
\hline \psi_{8} &0 & \frac{3}{4}J + V + 3\mu\\
\hline
\end{array}\\
\begin{array}{|c||c|c|}\hline
 & \psi_{9} & \psi_{10}\\
\hhline{|=#=|=|} \psi_{9} & \frac{3}{4}J + V + 3\mu  &
-\sqrt{2}t\\
\hline  \psi_{10} &-\sqrt{2} t  & 2\mu \\
\hline
\end{array}\quad
\begin{array}{|c||c|c|}\hline
 & \psi_{11} & \psi_{12}\\
\hhline{|=#=|=|} \psi_{11} & 2\mu &
-\sqrt{2} t\\
\hline \psi_{12} & -\sqrt{2}t & -\frac{1}{4}J + V + 3\mu\\
\hline
\end{array}\\
\begin{array}{|c||c|c|}\hline
 & \psi_{13} & \psi_{14}\\
\hhline{|=#=|=|} \psi_{13} & -\frac{1}{4}J + V + 3\mu &
-\sqrt{2} t\\
\hline \psi_{14} & -\sqrt{2}t & 2\mu\\
\hline
\end{array}\\
\begin{array}{|c||c|c|}\hline
 & \psi_{17} &\psi_{18}\\
\hhline{|=#=|=|} \psi_{17} &-\frac{1}{4}J + V + 3\mu & 0\\
\hline \psi_{18} &0 & -\frac{1}{4}J + V + 3\mu\\
\hline
\end{array}\\
\begin{array}{|c||c|}\hline
 & \psi_{20}\\
\hhline{|=#=|} \psi_{20} & J + 2V + 4\mu \\ \hline
\end{array}\quad
\begin{array}{|c||c|}\hline
 & \psi_{22}\\
\hhline{|=#=|} \psi_{22} & 2V + 4\mu \\ \hline
\end{array}\\
\begin{array}{|c||c|}\hline
 & \psi_{24}\\
\hhline{|=#=|} \psi_{24} & -\frac{1}{2}J +2V +4\mu \\ \hline
\end{array}\quad
\begin{array}{|c||c|}\hline
 & \psi_{25}\\
\hhline{|=#=|} \psi_{25} & -\frac{1}{2}J +2V +4\mu \\ \hline
\end{array}
\end{gather*}
\caption{Diagonal matrix blocks of the unrenormalized three-site
Hamiltonian $-\beta H(i,j)-\beta H(j,k)$.  The Hamiltonian being
invariant under spin-reversal, the spin-flipped matrix elements are
not shown.}
\end{table}
\endgroup

\appendix
\section{Derivation of the Decimation Relations}

The derivation of Eq. \eqref{eq:6}, first done in
Ref.\cite{FalicovBerker}, is given in this Appendix. In Eq.
\eqref{eq:5} the operators $-\beta^\prime H^\prime(i, k)$ and
$-\beta H(i, j)-\beta H(j, k)$ act on two-site and three-site
states, respectively, where at each site an electron may be either
with spin $\sigma=\uparrow$ or $\downarrow$, or may not exist ($0$
state). In terms of matrix elements,
\begin{multline}
\langle u_{i}v_{k}|e^{-\beta ^{\prime }H^{\prime }(i,k)}|\bar{u}_{i}^{{}}%
\bar{v}_{k}^{{}}\rangle = \label{eq:app1}\\
\sum_{w_{j}}\langle u_{i}\,w_{j}\,v_{k}|e^{-\beta H(i,j)-\beta
H(j,k)}|\bar{u}_{i}\,w_{j}\,\bar{v}_{k}^{{}}\rangle \:,
\end{multline}
where $u_{i},w_{j},v_{k},\bar{u}_{i},\bar{v}_{k}^{{}}$ are
single-site state variables, so that the left-hand side reflects a
$9\times9$ and the right-hand side a $27\times27$ matrix.  Basis
states that are simultaneous eigenstates of total particle number
($n$), parity ($p$), total spin magnitude ($s$), and total spin
z-component ($m_s$) block-diagonalize Eq. \eqref{eq:app1} and
thereby make it manageable.  These sets of 9 two-site and 27
three-site eigenstates, denoted by $\{|\phi _{p}\rangle \}$ and
$\{|\psi _{q}\rangle \}$ respectively, are given in Tables III and
IV.  Eq. \eqref{eq:app1} is thus rewritten as
\begin{multline}
\langle \phi _{p}|e^{-\beta ^{\prime }H^{\prime }(i,k)}|\phi _{\bar{p}%
}\rangle = \label{eq:app2}\\
\sum_{\substack{u,v,\bar{u},\\ \bar{v},w}}
\sum_{\substack{q,\bar{q}}} \langle\phi _p|u_iv_k\rangle \langle
u_iw_jv_k|\psi_q\rangle \langle \psi _q|e^{-\beta H(i,j)-\beta
H(j,k)}|\psi _{\bar{q}}\rangle\cdot \\
\langle \psi_{\bar{q}}|\bar{u}_iw_j\bar{v}_k\rangle \langle
\bar{u}_i\bar{v}_k|\phi _{\bar{p}}\rangle\:.
\end{multline}
There are five independent elements for $\langle \phi _{p}|e^{-\beta
^{\prime }H^{\prime }(i,k)}|\phi_{\bar{p}}\rangle$ in
Eq.\eqref{eq:app2} (thereby leading to five renormalized interaction
constants $\{t^\prime, J^\prime, V^\prime, \mu^\prime, G^\prime\}$),
which we label $\gamma_p$,
\begin{equation}\label{eq:app3}
\begin{split}
\gamma_p &\equiv \langle \phi _{p}|e^{-\beta ^{\prime }H^{\prime
}(i,k)}|\phi_{p}\rangle \quad \text{for}\: p = 1,2,4,6,7\,.
\end{split}
\end{equation}
The diagonal matrix $\langle \phi _{p}|-\beta ^{\prime }H^{\prime
}(i,k)|\phi_{\bar{p}}\rangle$ is given in Table V.  The exponential
of this matrix yields the five renormalized interaction constants in
terms of $\gamma_p$, as given in Eq. \eqref{eq:6}. Furthermore,
according to Eq. \eqref{eq:app2}, each $\gamma_p$ is a linear
combination of some $\langle \psi _q|e^{-\beta H(i,j)-\beta
H(j,k)}|\psi _{\bar{q}}\rangle$,
\begin{align*}\label{eq:app4}
\gamma_1&=\langle\psi_{1}||\psi_{1}\rangle\rsp+ \langle\psi_{2}||\psi_{2}\rangle\rsp+\langle\psi_{4}||\psi_{4}\rangle\,,\displaybreak[0]\\[5pt]
\gamma_2&=\langle\psi_{3}||\psi_{3}\rangle\rsp+ \frac{1}{2}\langle\psi_{8}||\psi_{8}\rangle\rsp+\langle\psi_{12}||\psi_{12}\rangle\rsp+\frac{1}{2}\langle\psi_{13}||\psi_{13}\rangle\,,\displaybreak[0]\\[5pt]
\gamma_4&=\langle\psi_{6}||\psi_{6}\rangle\rsp+ \frac{1}{2}\langle\psi_{9}||\psi_{9}\rangle\rsp+\langle\psi_{17}||\psi_{17}\rangle\rsp+\frac{1}{2}\langle\psi_{18}||\psi_{18}\rangle\,,\displaybreak[0]\\[5pt]
\gamma_6&=\langle\psi_{10}||\psi_{10}\rangle\rsp+2\langle\psi_{22}||\psi_{22}\rangle\,,\displaybreak[0]\\[5pt]
\gamma_7&=\langle\psi_{11}||\psi_{11}\rangle\rsp+\frac{2}{3}\langle\psi_{20}||\psi_{20}\rangle\rsp+\frac{4}{3}\langle\psi_{24}||\psi_{24}\rangle\,,\displaybreak[0]\\[5pt]
\end{align*}
where $\langle\psi_{q}||\psi_{q}\rangle\equiv\langle \psi
_{q}|e^{-\beta H(i,j)-\beta H(j,k)}|\psi_{q}\rangle$. In order to
calculate $\langle \psi _{q}|e^{-\beta H(i,j)-\beta
H(j,k)}|\psi_{\bar{q}}\rangle$ the matrix blocks in Table VI are
numerically exponentiated.

\end{document}